\documentclass[aps,prl,twocolumn,groupedaddress]{revtex4}
\usepackage{graphicx}
\begin{document}

\title{Experimental Polarization State Tomography using Optimal Polarimeters}

\author{Alexander Ling}
\email[]{phylej@nus.edu.sg}
\homepage[]{http://www.quantumlah.org}
\author{Soh Kee Pang}
\author{Ant\'{\i}a Lamas-Linares}
\author{Christian Kurtsiefer}
\affiliation{Department of Physics, National University of Singapore, Singapore, 117542}

\date{\today}

\begin{abstract}
We report on the experimental implementation of a polarimeter based on a scheme known to be optimal for obtaining the polarization vector of ensembles of $\mbox{spin-}\frac{1}{2}$ quantum systems, and the alignment procedure for this polarimeter. 
We also show how to use this polarimeter to estimate the polarization state for identically prepared ensembles of single photons and photon pairs and extend the method to obtain the density matrix for generic multi-photon states.  State reconstruction and performance of the polarimeter is illustrated by actual measurements on identically prepared ensembles of single photons and polarization entangled photon pairs. 
\end{abstract}


\maketitle
\section{I. Introduction}
Many promising applications in Quantum Information (e.g. Quantum Computation and Quantum Communication) demand accurate state estimation.  For many of them it is compelling to implement state estimation techniques that are both fast and consume as few copies of the state as possible.  Research in improving the efficiency of quantum state estimation techniques is an area of active theoretical study \cite{massar,derka,latorre,gill,schack,mixed1,mqt} with much focus on spin-$\frac{1}{2}$ systems (qubits).  Experimental reports on state estimation are fewer \cite{awhite}, partly because many schemes call for a joint measurement on an ensemble of qubits which is not always possible to implement.  

One way to realise qubits experimentally is to use single photons and consider their polarization degree of freedom, as this can be described by a two dimensional Hilbert space.  Estimating the polarization state of an ensemble of single photons (called polarimetry in classical optics) is equivalent to estimating the state of the qubit ensemble.  This makes efficient polarimetry interesting to Quantum Information.  

Polarimetry that uses the least number of measurement outcomes is said to be minimal.  Minimal polarimetry techniques in classical optics have been known for a long time and a lot of work in their optimization has been done \cite{handbook2,ambi1,ambi2,doafdp,sabatke1}.  While these classical methods perform well in estimating the polarization state for single photon ensembles in the limit of large numbers, their performance in the regime of extremely low light intensity (single photon level) was uncertain and it was not obvious how to use them in estimating non-classical states of light.  For this reason, progress in polarimetry at the single photon limit will assist in many areas, including characterization of faint sources of light, classical ellipsometry \cite{handbook2}, advanced quantum key distribution protocols \cite{tomo2,tomo1,sprot} and studies of the fundamental aspects of quantum theory \cite{aspect}.  

In discriminating between different estimation techniques we distinguish between methods that are minimal and those that are minimal {\it and} optimal \cite{latorre,mqt}: optimal methods have the best asymptotic efficiency in determining an unknown state when averaged over all possible input states.  This can be used for an operational definition of minimal and optimal state estimation for ensembles of prepared quantum systems. It is the technique that provides the best improvement to our estimated state for each additonal copy taken from the ensemble.  Recently \u{R}eh\'{a}\u{c}ek {\it et al.} proposed such a method for state estimation of polarization based single qubits \cite{mqt}, which can be viewed as an extension of classical techniques \cite{anafdp,ambi1,ambi2} to quantum systems.  

In this paper we address the experimental problem of implementing the optimal state estimation method described in \cite{mqt} by using a complete four output polarimeter with no moving parts.   
We describe that polarimeter in section II by reviewing the theory of optimal polarization state estimation and explain our implementation.  In section III polarization state estimation of multi-photon states is addressed.  In section IV we elaborate on the alignment procedure to make the polarimeter perform optimally.  Experimental state reconstruction on ensembles of single photons and photon pairs with high fidelity will be illustrated in sections V and VI. 

\section{II. State Estimation using the Optimal Polarimeter}

The polarisation state of light can be characterized using the three Stokes parameters $S_1,S_2,S_3$, possibly augmented by an intensity $S_m$.  Together they form a Stokes vector $\vec{S}$ and, when normalized, it is written as $\vec{S}=(1,S_1/S_m,S_2/S_m,S_3/S_m)$.  A {\it reduced} Stokes vector $\vec{S_r}=(S_1,S_2,S_3)/S_m$ identifies a point in the Poincare sphere (in this paper reduced Stokes vectors are denoted by an $r$ subscript) \cite{bloch}.  

\begin{figure} \scalebox{0.33}{\includegraphics{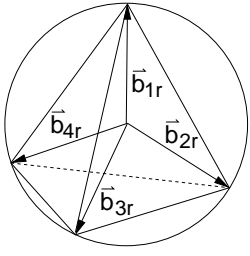}} \caption{ \label{fig:tetinsphere} Four (reduced) Stokes vectors in the Poincare sphere that form a tetrahedron define the optimal POVM operators used for polarization state estimation.  The tetrahedron gives the largest volume encompassable by a vector quartet in the sphere making it the optimal measurement when using four POVMs.} \end{figure}

A minimal scheme of estimating the Stokes vector requires exactly four detector readings, which corresponds to finding the overlap of the unknown Stokes vector with four non-coplanar vectors that define a tetrahedron in the Poincare sphere (Figure \ref{fig:tetinsphere}).  These four non-coplanar vectors define four measurement operators $B_j$ that govern the detector readings and form a set of complete Positive Operator Value Measurements (POVM) \cite{kraus}.  The tetrahedron geometry defines the largest volume that can be enclosed by a vector quartet in the Poincare sphere, making it the optimal estimation technique when using four POVMs \cite{ambi2,sle}.  
Such a state estimation technique is also unbiased in the asymptotic limit because the total distance of any vector in the Poincare sphere to all four POVM vectors depends only on the vector's magnitude.  In other words, the orientation of the unknown vector does not affect the accuracy with which it is estimated \cite{mqt}.  

We shall denote the tetrahedron's reduced Stokes vectors by $\vec{b_{1r}},\vec{b_{2r}},\vec{b_{3r}},\vec{b_{4r}}$ as shown in Figure \ref{fig:tetinsphere} and write their corresponding normalized vectors as $\vec{b_{1}},\vec{b_{2}},\vec{b_{3}},\vec{b_{4}}$.  Each measurement operator $B_j$ may be expressed as
\begin{eqnarray}
\label{eq:povmoperator}
B_j = \frac{1}{4}(\vec{b_{j}}\cdot\vec{\sigma}),
\end{eqnarray}
 where $\vec{\sigma} = (\sigma_0,\sigma_1,\sigma_2,\sigma_3)$,  $\sigma_0$ being the unit matrix and $\sigma_{1,2,3}$ the Pauli matrices.

In an experiment we associate each operator $B_j$ with a detector $\mbox{b}_j$.  The average intensity falling on detector $\mbox{b}_j$ is denoted as $I_j$.  Thus expectation values of the tetrahedron operators are related to detected intensities as follows:
\begin{eqnarray}
\label{eq:intensityeqn}
\frac{I_j}{I_t}=\langle B_j \rangle=\frac{1}{4}(\vec{b_j}\cdot\vec{S}) \qquad \mbox{ with } \qquad I_t=\sum^4_{j=1} I_j
\end{eqnarray}

Writing the intensities as a vector $\vec{I}=(I_1,I_2,I_3,I_4)/I_t$ gives us the Stokes vector 
\begin{eqnarray}
\label{eq:instresponse}
\vec{I} &=& \Pi\cdot\vec{S} \quad \Leftrightarrow \quad \vec{S}=\Pi^{-1}\cdot\vec{I} , 
\end{eqnarray} where $\Pi$ is referred to as the instrument matrix. Each row of this matrix is composed from a vector $\vec{b_j}$.  
One possible instrument matrix of the ideal polarimeter is:
\begin{eqnarray}
\label{eq:idealinstmatrix}
\Pi=\frac{1}{4}
\left(
\begin{array}{cccc}
1 & \sqrt\frac{1}{3} & \sqrt\frac{2}{3} & 0 \\
1 & \sqrt\frac{1}{3} & -\sqrt\frac{2}{3} & 0 \\
1 & -\sqrt\frac{1}{3} & 0 & -\sqrt\frac{2}{3} \\
1 & -\sqrt\frac{1}{3} & 0 & \sqrt\frac{2}{3}
\end{array}
\right)
\end{eqnarray}

\begin{figure} 
\scalebox{0.48}{\includegraphics{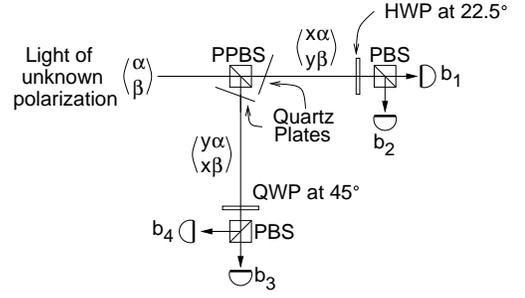}} 
\caption{
\label{fig:polarimeterscheme}
Practical implementation of the tetrahedron polarimeter that achieves the ideal instrument matrix.  Each detector $\mbox{b}_j$ is associated to the tetrahedron vector $\vec{b_j}$.  The partially polarizing beam splitter (PPBS) separates incoming light according to polarization, and quartz plates remove unwanted phase shifts.  Light leaving the PPBS is passed through waveplates and polarizing beam splitters (PBS) to be projected on two different bases ($\pm 45^{\circ}$ basis for transmitted light and the circular basis for reflected light).} \end{figure}

Experimental realisation of this instrument matrix is achieved by the polarimeter shown in Figure \ref{fig:polarimeterscheme}.  The first component of the polarimeter is a partially polarising beam splitter (PPBS) that has a particular amplitude splitting ratio for incoming light, most easily determined using Jones vector notation for polarization.  The amplitude division coefficients of the PPBS $x$ and $y$ obey energy conservation  $|x^2| + |y^2| = 1$.  The PPBS takes horizontally polarized (H) light $1\choose 0$ to the polarizations $x\choose 0$ and $y\choose 0$ in the transmitted and reflected arm, respectively, and vertically (V) polarized light $0\choose 1$ to $0\choose y$ in transmission and $0\choose x$ in reflection.  

If we project light in the transmitted arm of the PPBS on the $\pm 45^{\circ}$ polarization basis and light in the reflected arm onto the Left/Right circular polarization basis, the tetrahedral arrangement of the vectors $\vec{b_j}$ are ensured with the following relations:
\begin{eqnarray}
\label{tetrahedronrel}
x^2=\frac{1}{2}+\frac{1}{2\sqrt{3}}\mbox{\quad and}\quad y^2=\frac{1}{2}-\frac{1}{2\sqrt{3}}.\end{eqnarray}  
Detailed steps are given in Appendix A.

Partially polarized light can be described using a density matrix (or coherency matrix).  If we write the entries of the density matrix as a column vector $\vec{\rho}$, we can determine them from the Stokes vector \cite{statopt} using the following transformation:
\begin{eqnarray}
\label{eq:stokestransform}
\vec{\rho}&=&\frac{1}{2}\Gamma_1\cdot \vec{S}\nonumber\\
&=&\frac{1}{2}\left( 
\begin{array}{cccc}
1 & 1 & 0 & 0\\
0 & 0 & 1 & i\\
0 & 0 & 1 & -i\\
1 & -1 &0 & 0
\end{array}
\right)
\cdot\vec{S}
\end{eqnarray}

The columns of the matrix $\Gamma_1$ are the Pauli operators written as column vectors
$\Gamma_1=(\vec{\sigma_0},\vec{\sigma_1},\vec{\sigma_2},\vec{\sigma_3})$.  
The matrices $B^{-1}$ and $\Gamma_1$ can be combined into a single matrix \begin{eqnarray} \label{tomomatrix} T:=\frac{1}{2}\Gamma_1 \Pi^{-1} \quad \Rightarrow \quad \vec{\rho}=T\cdot\vec{I}\end{eqnarray}
 which might be referred to as a tomography matrix as it directly relates the detected intensities to the density matrix of the state.

\section{III. Polarization State Tomography for Ensembles with multi-photons}
The instrument matrix scheme above can be extended to perform polarization state tomography on ensembles of multi-photon states.  James {\it et al.} \cite{awhite} have described a similar state estimation method.  We follow their approach but use our optimal and instrumentally motivated measurement operators, thereby reducing any ambiguity over the choice of operators.

\begin{figure}
\scalebox{0.17}{\includegraphics{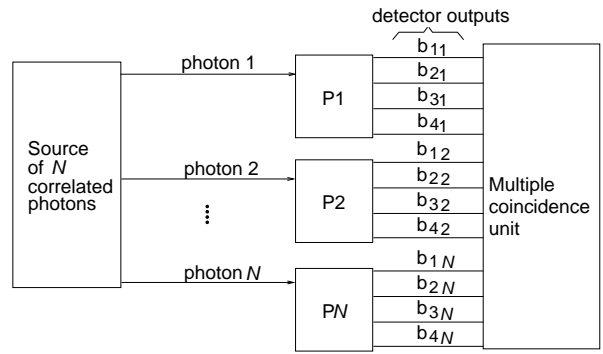}}
\caption{\label{fig:multiphoton} Scheme for estimating the polarization state of an ensemble of $N$-correlated photons using $N$ polarimeters (P1,P2,...,P$N$).  A multiple coincidence circuit identifies the $4^N$ possible coincidence combinations.  For photon pairs ($N$=2), two polarimeters are used giving 16 possible coincidence combinations.  Several copies of the state are processed giving a coincidence pattern used in estimating the polarization state of the ensemble.}
\end{figure}

The simplest multi-photon system is a photon pair detected by testing for coincidence in the detection time of their component photons.  In our measurement process each member of the photon pair  is passed through a polarimeter.  
Given two polarimeters $1$ and $2$, each with four detectors $\mbox{b}_{i_1} \mbox{ and } \mbox{b}_{i_2}$, respectively, ($i_{1},i_{2}=0,1,2,3$), we will have 16 possible coincidence combinations between the detectors (see Figure \ref{fig:multiphoton}).  Each coincidence rate is governed by an operator composed from the individual detectors' measurement operators.  
If we denote again the measurement operator of detectors $\mbox{b}_{i_1} \mbox{ and }\mbox{b}_{i_2}$ as $B_{i_1}$ and $B_{i_2}$, and the coincidence count between them as $c_{i_1,i_2}$, we can express the coincidence rates as a linear function of a 2-photon polarization state vector $S_2$:
\begin{eqnarray}
\label{eq:biphotonstokes}
\frac{c_{i_1,i_2}}{c_{t}} = \langle B_{i_1} \otimes B_{i_2}\rangle
&=& (\frac{1}{4}\vec{b_{i_1}}\otimes \frac{1}{4}\vec{b_{i_2}})\cdot \vec{S_2},\\
\mbox{with } c_t&=&\sum_{i_1,i_2=1}^{4} c_{i_1,i_2} \nonumber
\end{eqnarray} 
Here, $\vec{S_2}$ is the Stokes vector equivalent for a 2-photon system \cite{awhite} and $c_t$ is the total number of observed coincidences.  We now have the set of measurement operators governing the coincidence pattern.  The sixteen coincidences $c_{i_1,i_2}$ can be written in column vector format $\vec{C_2}=$($c_{1,1},c_{1,2},...,c_{4,4})$.  If we define the 2-polarimeter instrument matrix as $\Pi_2$, we obtain an instrument response analogous to (3):  
\begin{eqnarray}
\label{eq:biphotoninstmatrix}
\vec{C_2} = \Pi_2 \cdot \vec{S_2} \Leftrightarrow 
\vec{S_2}= \Pi^{-1}_2 \cdot \vec{C_2}
\end{eqnarray}

Thus we obtain the density matrix of the 2-photon state by constructing the analogous 2-photon expression for equation (\ref{eq:stokestransform}):
\begin{eqnarray} 
\label{eq:biphotontransform}
\vec{\rho_2}= \frac{1}{2^2}\Gamma_2 \cdot \vec{S_2} = T_2 \cdot\vec{C_2}\end{eqnarray}
Each column of $\Gamma_2$ is the product of two Pauli operators $\sigma_{i_1}\otimes\sigma_{i_2}$ ($i_1,i_2=0,1,2,3$) written in column vector format and $T_2$ is the tomography matrix for the 2-photon state.  

It is now straightforward to generalize this concept to obtain the density matrix for states of $N$ correlated photons.  Using $N$ polarimeters, we obtain the pattern of $N$-fold coincidences to build up the coincidence vector $\vec{C_N}$ which is used to find the $N$-photon Stokes vector and density matrix:
\begin{eqnarray}
\label{eq:generalstokes}
\vec{S_N} &=& \Pi_N^{-1}\cdot\vec{C_N},\\
\vec{\rho_N}&=& \frac{1}{2^N}\Gamma_N \cdot \vec{S_N} = T_N \cdot \vec{C_N} 
\end{eqnarray}
Each row of the instrument matrix $\Pi_N$ is given by $(\frac{1}{4}\vec{b_{i_1}}\otimes\frac{1}{4}\vec{b_{i_2}}...\otimes\frac{1}{4}\vec{b_{i_N}})$ and
 each column of $\Gamma_N$ is the product of $N$ Pauli matrices $\sigma_{i_1}\otimes\sigma_{i_2}...\otimes\sigma_{i_n}$ ($i_n = 0,1,2,3$ \mbox { and } $n=1,2,...,N$).  This generalized approach will work for all four-detector polarimeters in multi-photon analysis schemes (Figure \ref{fig:multiphoton}).  We note that --- although the polarimeters are optimal for estimating single qubit states --- it is an open question if the scheme above is optimal in estimating multi-photon systems. 

\begin{figure*} 
\scalebox{0.27}{\includegraphics{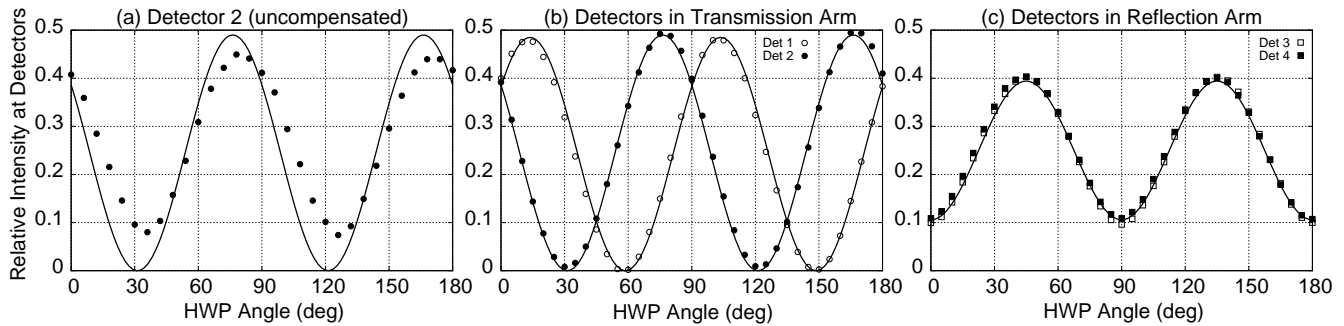}}
\caption{
\label{fig:instresponse}
Instrument response of the polarimeter to linearly polarized light.  The data points show the variation in relative intensity at each detector with respect to the angle of the half wave plate (HWP) in the polarisation state preparation.  The solid lines show the expected intensity modulation for an ideal device for each HWP setting (\ref{eq:linearpolresponse}), scaled for appropriate detector efficiencies.  Error bars are smaller than the point markers.  Panel (a) shows the relative intensity at detector 2 without compensation plates.  Panels (b) and (c) are taken with compensation for phase shifts.  The oscillation in (a) is out of phase and also of lower amplitude compared to the phase corrected behaviour of detector 2 in panel (b).  Plots in (c) shows a lower amplitude because light in the reflected arm is not projected on a linear polarization basis.} \end{figure*} 

\section{IV. Phase Correction and Polarimeter Calibration} 

\subsection{IV a. Removing Unwanted Phase Shifts} 
In the presented polarimeter, an ideal PPBS has the nominal beam splitting ratio (\ref{tetrahedronrel}) and also rotates the polarization state of light leaving the beamsplitter into the correct polarization basis \cite{doafdp}. Such beamsplitters, however, are not easily available and their design is the focus of active research \cite{doafdpbs}.  We therefore use beamsplitters with only the nominal intensity splitting ratio.  

A PPBS without phase shift diverts light in state $-\vec{b_{jr}}$ (that is conjugate to a tetrahedron vector $\vec{b_{jr}}$) from detector $\mbox{b}_j$.  General beamsplitters, however, lack this phase preserving property.  The result is that input of conjugate states $\vec{-b_{jr}}$ does not stop light from reaching the associated detectors.  This suggests an easy alignment method for correcting any unwanted phase shifts with birefringent compensation plates.  

For phase correction we first prepare high quality H-polarized light using polarizers of extinction ratio $10^5$.  With one subsequent half wave plate (HWP) and one quarter wave plate (QWP) we can then prepare any polarization state on the surface of the Poincare sphere.  Compensator plates (0.5mm thick quartz) mounted on rotating stages were placed at each output arm of the PPBS, and light with a conjugate polarization state was sent to the polarimeter.  For each polarization state $-\vec{b_{jr}}$ the compensator in the relevant output arm was rotated until the detector $\mbox{b}_j$ received no light.  Two input states (one for each output arm) were sufficient to compensate for the unwanted phase shifts.

We verified the compensated polarimeter behavior with linearly polarized light prepared using only the polarizer and HWP (this reduces preparation errors due to residual errors in the QWP).  The prepared states have a Stokes vector of the form (1,$\cos4\psi$,$\sin4\psi$,0), where $\psi$ is the angle of the HWP, so the normalized response of detector 1, for example, will be  
\begin{eqnarray} 
\label{eq:linearpolresponse}
I_1 = 1 + \sqrt{\frac{1}{3}}\cos4\psi + \sqrt{\frac{2}{3}}\sin4\psi. \end{eqnarray} 

Passively quenched silicon avalanche photodiodes were used as detectors allowing us to perform photon counting.  
The number of photons accumulated at each detector output was noted for each angle of the HWP.  The results are shown in Figure \ref{fig:instresponse}.  

The results show that the response of the compensated polarimeter is very close to ideal. The extrema of our measured intensities are less than $1^{\circ}$ (of HWP angle) away from their nominal positions.  This means that our actual measurement vectors are pointing in the same direction as the ideal tetrahedron vectors, although their magnitudes will be different due to imbalanced detection efficiencies.  While this renders the asymptotic efficiency of our polarimeter less than ideal, it still represents the optimal setup for the collection efficiencies we can achieve.  In other words, we are maximizing the volume defined by our experimental POVM vectors \cite{sle}.  

This measurement result is limited by the accuracy of our rotation controllers.  Our waveplates are mounted on rotary motors with an accuracy of $0.3^{\circ}$.  The polarizing beam splitters in the output arms have an extinction ratio of $10^4$ and the waveplates' optical path length differ from their nominal values by less than 2\%.
\subsection{IV b. Calibrating the Polarimeter} 
We calibrate the instrument matrix of this polarimeter to account for all residual phase shifts and coupling inefficiences.  A general calibration technique for four detector polarimeters (``equator-poles method") was described by Azzam {\it et al.} \cite{anafdp}.  Incidentally, the phase dependency measurement shown in Figure \ref{fig:instresponse} was an essential part of this calibration.  

Using this technique we are able to find the correction terms needed to be made to our ideal instrument matrix.  A typical corrected instrument matrix $\Pi_c$ is shown below: 
\begin{eqnarray*} \Pi_c=\frac{1}{4} \left( \begin{array}{cccc} 0.962 & 1.051\sqrt\frac{1}{3} & 0.920\sqrt\frac{2}{3} & 0.005 \\ 0.991 & 1.031\sqrt\frac{1}{3} & -0.956\sqrt\frac{2}{3} & -0.005 \\ 1.010 & -1.045\sqrt\frac{1}{3} & 0.005 & -0.945\sqrt\frac{2}{3} \\ 1.032 & -1.009\sqrt\frac{1}{3} & 0.029 & 1.011\sqrt\frac{2}{3} \end{array} \right). \end{eqnarray*} 
The uncertainty for each of the correction terms above is on the order of $0.002$.  We see that the deviation from entries in the ideal instrument matrix (\ref{eq:idealinstmatrix}) is on the order of a few percent.

The phase correction and calibration steps presented above must take into account the wavelength of the input light because optical elements are specified to perform only within a certain bandwidth.  The polarimeter was built to study the polarization state of light coming from a spontaneous parametric down conversion (SPDC) source \cite{brsrc} with a spectral bandwidth of $4.740\pm 0.014$ nm centred around 702nm.  The same light source was used for phase correction and polarimeter calibration and the experiments described in the remaining sections.

\section{V. Experimental State Tomography for Ensembles of Single Photons} 
\begin{figure} 
\scalebox{0.5}{\includegraphics{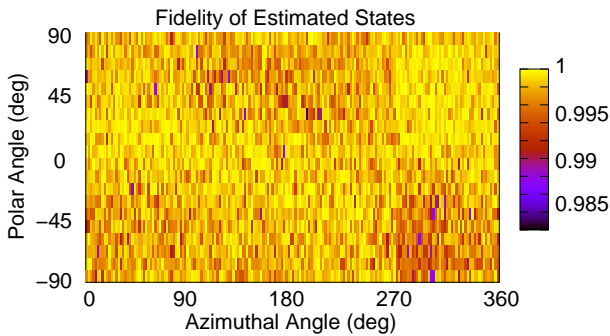}} 
\caption{
\label{fig:fidelitymap}
A set of polarization states ($\vec{S_i}$) equally distributed over the Poincare sphere surface was generated; photons from each of these states were sent to the polarimeter, from which an estimated state ($\vec{S_e}$) is obtained.  The figure shows the fidelity of the estimated state to the prepared state, $\frac{1}{2}(\vec{S_e}\cdot \vec{S_i})$.  It is  roughly constant over the Poincare sphere, showing that the polarimeter is an unbiased polarization state estimator.
} \end{figure}
The ability of the tetrahedron polarimeter to estimate polarization states without bias was tested by preparing a set of pure polarization states equally distributed over the Poincare sphere.  In this way we could better identify regions that suffer poor state estimation (if any).

Computer controlled motors were used to rotate waveplates (after a H-filter) in preparing the set of polarization states.  The Stokes vector of a pure polarization state can be expressed as $\vec{S} = (1,\cos2\delta \cos2[\psi+\delta],-\cos2[\psi+\delta]\sin2\delta,-\sin2[\psi+\delta])$, where $\delta \mbox{ and } \psi$ are the QWP and HWP angles, respectively.  Thus any set of coordinates (characterized by the polar and azimuthal angles) on the Poincare sphere can be expressed in terms of the waveplate angles.  

For each set of angles, the detectors accumulated photon detection events for one second giving a particular vector $\vec{I}$ from which an estimated Stokes vector $\vec{S_e}$ and probability density matrix $\rho_e$ can be obtained via equations (\ref{eq:instresponse}) and (\ref{eq:stokestransform}).  To calculate the distance of the estimated state from the (ideal) prepared state $\rho_i$ ($\vec{S_i}$), we use the Uhlmann fidelity, defined as $(\mbox{tr}[\sqrt{\sqrt{\rho_i}\rho_e\sqrt{\rho_i}}])^2$ \cite{uhlmann,jozsa}.  For pure states this quantity reduces to the overlap of their Stokes vectors $\frac{1}{2}(\vec{S_i}\cdot\vec{S_e})$.   

The fidelity was mapped to the appropriate polar and azimuthal coordinates on the Poincare sphere (Figure \ref{fig:fidelitymap}); linear polarization states correspond to a polar angle of $0^\circ$.  
The average fidelity for the whole map is 99.8\% with a minimum fidelity of 98.4$\pm$0.9\% (the cumulative photon count per point is approximately 2000).  There are no systematic areas of low fidelity even when wedge errors in the state preparation waveplates cause count rates to drop.  This indicates that the polarimeter estimates all pure polarization states equally well.

Fidelity does not distinguish between errors introduced in state preparation from errors in the state estimation process.  Therefore we have characterized our state preparation apparatus independently and are confident that their contribution to the error in calculated fidelity above is on the order of $\pm 0.01$\%.  Thus we assign the residual difference in fidelity to imperfections in the detection apparatus. 

\section{VI. Experimental state tomography for a two photon ensemble}
We will now illustrate how to use two polarimeters to perform polarization state tomography on a 2-photon state generated from an SPDC source.  First, two polarimeters were correctly aligned and after calibration their instrument matrices were found to be:
\begin{eqnarray*}
\frac{1}{4}
\left(
\begin{array}{cccc}
0.903 & 0.927\sqrt\frac{1}{3} & 0.9997\sqrt\frac{2}{3} & -0.041 \\
1.124 & 1.135\sqrt\frac{1}{3} & -1.014\sqrt\frac{2}{3} & 0.0602 \\
0.995 & -1.079\sqrt\frac{1}{3} & 0.001 & 0.913\sqrt\frac{2}{3} \\
0.978 & -0.983\sqrt\frac{1}{3} & 0.003 & -0.936\sqrt\frac{2}{3}
\end{array}
\right)
\end{eqnarray*}
and
\begin{eqnarray*}
\frac{1}{4}
\left(
\begin{array}{cccc}
1.074 & 1.171\sqrt\frac{1}{3} & 0.913\sqrt\frac{2}{3} & -0.082 \\
0.983 & 0.8804\sqrt\frac{1}{3} & -1.044\sqrt\frac{2}{3} & 0.004 \\
1.082 & -1.172\sqrt\frac{1}{3} & 0.001 & -0.9625\sqrt\frac{2}{3} \\
0.862 & -0.88\sqrt\frac{1}{3} & -0.002 & 0.867\sqrt\frac{2}{3}
\end{array}
\right).
\end{eqnarray*}

We then arranged for the SPDC source to generate photon pairs that are detected as a maximally entangled Bell state $|\Psi^+\rangle$.  
Bell states created via SPDC are typically characterized by a polarization correlation experiment, from which a visibility value can be obtained \cite{brsrc}.  The visibility measured in the HV and $\pm 45^\circ$ basis was above $97.7\pm2\%$; such a high value is usually taken as evidence of a high degree of entanglement.

The photon pairs were passed through the polarimeters and the pattern of coincidences between them  was observed.  The 16 observed coincidence rates (collected using the scheme similar to \cite{multichannel}) make up the coincidence vector $\vec{C}$ =(21444, 1505, 24104, 26002, 979, 24716, 23210, 22447, 21661, 30752, 24061, 268, 19010, 23692, 339, 17695).

\begin{figure} 
\scalebox{0.4}{\includegraphics{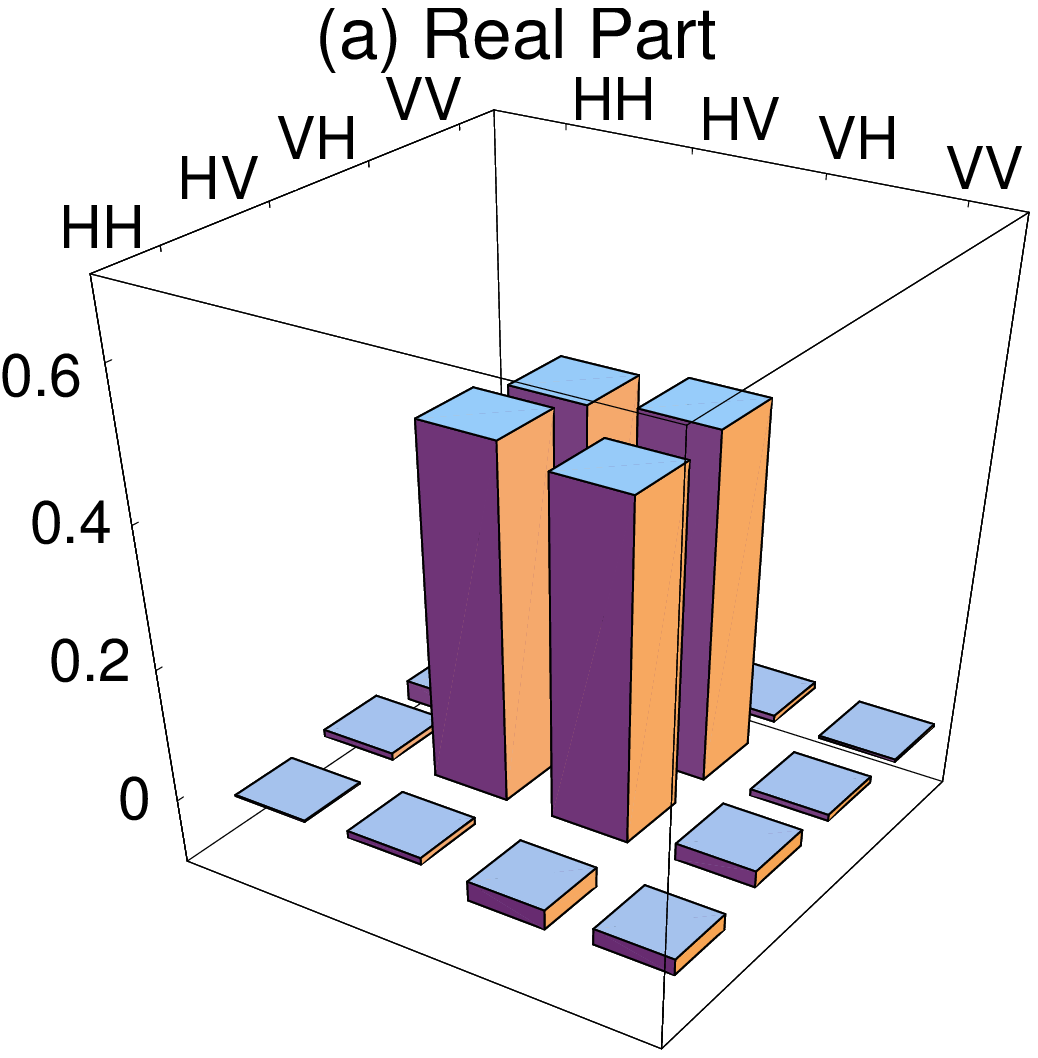}} \scalebox{0.4}{\includegraphics{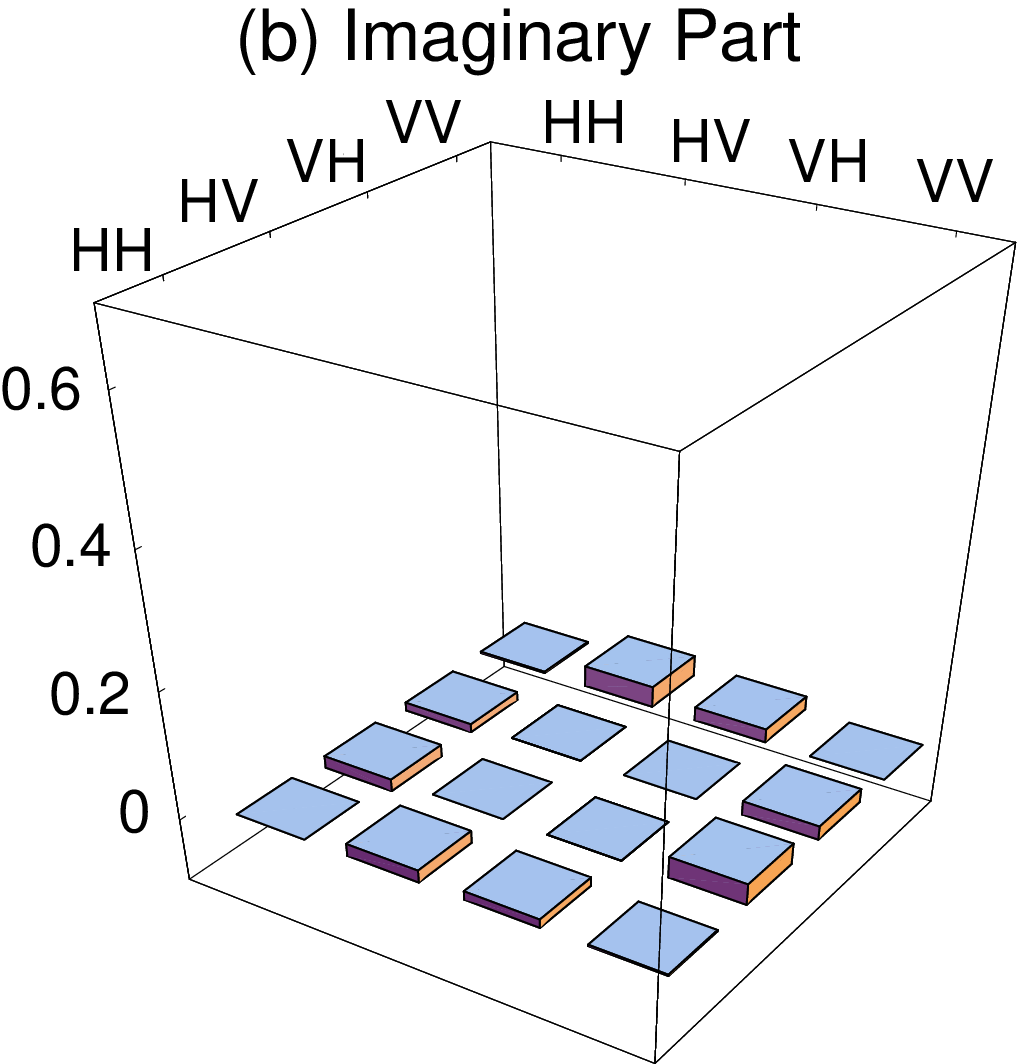}} 
\caption{
\label{fig:densitymatrix}
The density matrix of a Bell state $|\Psi^+\rangle\langle\Psi^+|$ obtained by linear reconstruction from photon pairs.} \end{figure}

Using this vector with equations (\ref{eq:biphotoninstmatrix}) and (\ref{eq:biphotontransform}) we obtain the density matrix whose real components are
\begin{eqnarray*}\mbox{Re}[\rho]=\left(\begin{array}{cccc}                 
-0.002 & -0.01 & -0.03 & -0.024 \\
-0.01 & 0.506 &  0.485 &  0.025 \\               
-0.03 & 0.485 &  0.498 &  0.009 \\
-0.024 & -0.024 &  0.009 & -0.003 
\end{array}                                                                     \right),                                                                         \end{eqnarray*}      
while the magnitude of the imaginary components are below a value of 0.04 (see Figure \ref{fig:densitymatrix}).  

The uncertainty in each of the above terms is on the order of 0.011.  The Uhlmann fidelity of this state to the ideal $|\Psi^+\rangle$ state was found to be 0.990 $\pm$ 0.014.  Error bars in all cases were computed by numerical derivation and propagated Poissonian counting noise. 
The propagated error bars result in an estimated density matrix compatible with the ideal $|\Psi^+\rangle \langle\Psi^+|$ state.

\section{VII. Conclusion}
In this paper we have illustrated a simple alignment procedure for optimizing the tetrahedron polarimeter.  Phase shifts introduced by a commercially available PPBS were easily corrected by phase compensation plates in the output arms.  The response of the compensated polarimeter was measured over a dense sampling of states on the Poincare sphere, and found to be similar to that of an ideal device.  This shows that beamsplitters need only have the nominal intensity splitting ratio, making optimal polarimeters more accessible.

We also described an instrumentally motivated method for constructing the measurement operators governing light distribution to each output of the polarimeter.  This instrument based approach also allows a convenient generalization to obtain measurement operators governing multi-photon coincidences.  These operators can then be applied to the linear reconstruction of multi-photon Stokes vectors and their density matrices.  

Optimal polarimeters were then used for estimating the polarization state of experimentally prepared ensembles of single photons and photon pairs in a Bell state.  The estimated states were evaluated by computing their fidelity to the (ideal) prepared states.  We found an average fidelity above 99.8\% in all our experiments.  Thus we have built and demonstrated the use of optimal 4-output polarimeters in multi-photon polarization state tomography.

While preparing this document, it came to our attention that a similar four output polarimeter was suggested independently in a recent paper \cite{aspect}.
\section{Acknowledgments}
\begin{acknowledgments}
We would like to thank Ivan Marcikic, Berge Englert and Janet Anders for helpful discussions.  This work was supported by DSTA Grant No. R-394-000-019-422 and A*Star Grant No. R-144-000-071-305.
\end{acknowledgments}

\section{Appendix A}
To derive the intensity splitting ratio of the PPBS, we first express the polarization states are expressed using Jones vectors unless we are describing the tetrahedron vectors $\vec{b_j}$.
The tetrahedron (Stokes) vectors $\vec{b_{j,k}}$ have the scalar product property 
\begin{eqnarray}
\label{eq:tetrahedronproperty}
\vec{b_j}\cdot\vec{b_k}=\frac{2}{3} + \frac{4}{3}\delta_{jk}.
\end{eqnarray}

Recalling the parameters of the intensity splitting ratio of the PPBS $x$ and $y$ we see that a general input polarization state $ \alpha \choose \beta$  leads to the polarizations $x\alpha \choose y\beta$ and $y\alpha \choose x\beta$ in the transmitted and reflected arms of the PPBS respectively.  In our polarimeter, light leaving the arms of the PPBS must be analyzed in two different polarization bases.  Two orthogonal vectors that form a basis may be expressed as $\cos\theta \choose e^{i\phi} \sin\theta$ and $-e^{-i\phi}\sin\theta \choose \cos\theta$.  This leads for example to the normalized light intensity falling on detector $\mbox{b}_1$  
\begin{eqnarray} 
\label{eq:lightondet1}
I_1/I_t &=&\left|\alpha x \cos\theta + \beta y e^{-i\phi} \sin\theta\right|^2.
\end{eqnarray} 

We choose a different measurement basis for detectors 3 and 4, for example light reaching detector 3 is 
\begin{eqnarray} 
\label{eq:lightondet3}
I_3/I_t &=& \left|\alpha y \cos\theta ' + \beta x e^{-i \phi '} \sin\theta '\right|^2.
\end{eqnarray}           

Using the vector $\vec{b_1}$ as an example, equation (\ref{eq:intensityeqn}) allows us to express the operator $B_1$ in terms of the measurement basis to fulfill equation (\ref{eq:lightondet1}):
\begin{eqnarray} 
\label{eq:newB1}
\langle B_1\rangle &=& \left|\alpha x \cos\theta + \beta y e^{-i\phi} \sin\theta\right|^2 \\ &=& \left| \left( \begin{array}{cc} x\cos\theta & y e^{-i\phi}\sin\theta \end{array}\right) {\alpha  \choose \beta }\right|^2\nonumber\\ 
\label{eq:B1breakdown}
\end{eqnarray}

The following choice of $B_1$ fulfills this condition
\begin{eqnarray}
B_1 &=& {x\cos\theta \choose y\sin\theta e^{i \phi}} \left ( x\cos\theta \quad y\sin\theta e^{-i\phi}\right ). \end{eqnarray}

Since the tetrahedron can be oriented arbitrarily we choose for convenience to measure the $45^{\circ}$ linear polarization basis ($\theta=\pi/4,\quad \phi=0$) in the transmitted arm and the circular polarization basis ($\theta '= \pi/4 \quad \phi '=\pi/2$) in the reflected arm.  This reduces the measurement operators to only the beamsplitting parameters $x$ and $y$
\begin{eqnarray*}
\left. \begin{array}{c}B_1\\B_2\end{array}\right\} = \frac{1}{2}{{x^2 \quad \pm xy}\choose {\pm xy \quad y^2}},\quad 
\left. \begin{array}{c}B_3\\B_4 \end{array}\right\} = \frac{1}{2}{{y^2 \quad \mp ixy}\choose {\pm ixy \quad x^2}},
\end{eqnarray*}
which together with equation (\ref{eq:intensityeqn}) allows us to express all tetrahedron vectors in terms of $x$ and $y$ 

\begin{eqnarray} 
\label{eq:tetrahedronvectors}
\left. \begin{array}{c} \vec{b_1} \\ \vec{b_2} \end{array} \right\}  =  \left( \begin{array}{c} 1 \\ x^2-y^2 \\ \pm 2xy \\0 \end{array} \right), \left. \begin{array}{c} \vec{b_3} \\ \vec{b_4} \end{array} \right\}  =  \left( \begin{array}{c} 1 \\ y^2-x^2 \\ 0 \\ \mp 2xy \end{array} \right). \end{eqnarray}

From equation (\ref{eq:tetrahedronproperty}), we can write: 
\begin{eqnarray}
\label{eq:vectormultiplication}
\vec{b_1}\cdot\vec{b_2}=\frac{2}{3} \mbox{ and }
\vec{b_1}\cdot\vec{b_3}=\frac{2}{3}.
\end{eqnarray}
This allows us to obtain an equation in $x$ alone
\begin{eqnarray}
\label{eq:quadraticeqn}
36x^8-24x^4+1&=&0.
\end{eqnarray}

The last equation gives two solution sets; we choose the set where 
$x^2=\frac{1}{2}+\frac{1}{2\sqrt{3}}\quad\Rightarrow\quad y^2=\frac{1}{2}-\frac{1}{2\sqrt{3}}$.

\bibliography{cwn}

\begin{thebibliography}{27}
\expandafter\ifx\csname natexlab\endcsname\relax\def\natexlab#1{#1}\fi
\expandafter\ifx\csname bibnamefont\endcsname\relax
  \def\bibnamefont#1{#1}\fi
\expandafter\ifx\csname bibfnamefont\endcsname\relax
  \def\bibfnamefont#1{#1}\fi
\expandafter\ifx\csname citenamefont\endcsname\relax
  \def\citenamefont#1{#1}\fi
\expandafter\ifx\csname url\endcsname\relax
  \def\url#1{\texttt{#1}}\fi
\expandafter\ifx\csname urlprefix\endcsname\relax\def\urlprefix{URL }\fi
\providecommand{\bibinfo}[2]{#2}
\providecommand{\eprint}[2][]{\url{#2}}

\bibitem[{\citenamefont{Massar and Popescu}(1995)}]{massar}
\bibinfo{author}{\bibfnamefont{S.}~\bibnamefont{Massar}} \bibnamefont{and}
  \bibinfo{author}{\bibfnamefont{S.}~\bibnamefont{Popescu}},
  \bibinfo{journal}{Phys. Rev. Lett.} \textbf{\bibinfo{volume}{74 (8)}},
  \bibinfo{pages}{1259} (\bibinfo{year}{1995}).

\bibitem[{\citenamefont{Derka et~al.}(1998)\citenamefont{Derka, Bu{\u{z}}ek,
  and Ekert}}]{derka}
\bibinfo{author}{\bibfnamefont{R.}~\bibnamefont{Derka}},
  \bibinfo{author}{\bibfnamefont{V.}~\bibnamefont{Bu{\u{z}}ek}},
  \bibnamefont{and} \bibinfo{author}{\bibfnamefont{A.~K.} \bibnamefont{Ekert}},
  \bibinfo{journal}{Phys. Rev. Lett.} \textbf{\bibinfo{volume}{80 (8)}},
  \bibinfo{pages}{1571} (\bibinfo{year}{1998}).

\bibitem[{\citenamefont{Latorre et~al.}(1998)\citenamefont{Latorre, Pascual,
  and Tarrach}}]{latorre}
\bibinfo{author}{\bibfnamefont{J.~I.} \bibnamefont{Latorre}},
  \bibinfo{author}{\bibfnamefont{P.}~\bibnamefont{Pascual}}, \bibnamefont{and}
  \bibinfo{author}{\bibfnamefont{R.}~\bibnamefont{Tarrach}},
  \bibinfo{journal}{Phys. Rev. Lett.} \textbf{\bibinfo{volume}{81 (7)}},
  \bibinfo{pages}{1351} (\bibinfo{year}{1998}).

\bibitem[{\citenamefont{Gill and Massar}(2000)}]{gill}
\bibinfo{author}{\bibfnamefont{R.~D.} \bibnamefont{Gill}} \bibnamefont{and}
  \bibinfo{author}{\bibfnamefont{S.}~\bibnamefont{Massar}},
  \bibinfo{journal}{Phys. Rev. A} \textbf{\bibinfo{volume}{61}},
  \bibinfo{pages}{042312} (\bibinfo{year}{2000}).

\bibitem[{\citenamefont{Schack et~al.}(2000)\citenamefont{Schack, Brun, and
  Caves}}]{schack}
\bibinfo{author}{\bibfnamefont{R.}~\bibnamefont{Schack}},
  \bibinfo{author}{\bibfnamefont{T.~A.} \bibnamefont{Brun}}, \bibnamefont{and}
  \bibinfo{author}{\bibfnamefont{C.~M.} \bibnamefont{Caves}},
  \bibinfo{journal}{Phys. Rev. A} \textbf{\bibinfo{volume}{64}},
  \bibinfo{pages}{014305} (\bibinfo{year}{2000}).

\bibitem[{\citenamefont{Bagan et~al.}(2004)\citenamefont{Bagan, Baig,
  Mu{\~n}oz-Tapia, and Rodriguez}}]{mixed1}
\bibinfo{author}{\bibfnamefont{E.}~\bibnamefont{Bagan}},
  \bibinfo{author}{\bibfnamefont{M.}~\bibnamefont{Baig}},
  \bibinfo{author}{\bibfnamefont{R.}~\bibnamefont{Mu{\~n}oz-Tapia}},
  \bibnamefont{and}
  \bibinfo{author}{\bibfnamefont{A.}~\bibnamefont{Rodriguez}},
  \bibinfo{journal}{Phys. Rev. A} \textbf{\bibinfo{volume}{69}},
  \bibinfo{pages}{010304(R)} (\bibinfo{year}{2004}).

\bibitem[{\citenamefont{\u{R}eh\'{a}\u{c}ek
  et~al.}(2004)\citenamefont{\u{R}eh\'{a}\u{c}ek, Englert, and
  Kaszlikowski}}]{mqt}
\bibinfo{author}{\bibfnamefont{J.}~\bibnamefont{\u{R}eh\'{a}\u{c}ek}},
  \bibinfo{author}{\bibfnamefont{B.-G.} \bibnamefont{Englert}},
  \bibnamefont{and}
  \bibinfo{author}{\bibfnamefont{D.}~\bibnamefont{Kaszlikowski}},
  \bibinfo{journal}{Phys.\ Rev.\ A} \textbf{\bibinfo{volume}{70}},
  \bibinfo{pages}{052321} (\bibinfo{year}{2004}).

\bibitem[{\citenamefont{James et~al.}(2001)\citenamefont{James, Kwiat, Munro,
  and White}}]{awhite}
\bibinfo{author}{\bibfnamefont{D.}~\bibnamefont{James}},
  \bibinfo{author}{\bibfnamefont{P.~G.} \bibnamefont{Kwiat}},
  \bibinfo{author}{\bibfnamefont{W.~J.} \bibnamefont{Munro}}, \bibnamefont{and}
  \bibinfo{author}{\bibfnamefont{A.~G.} \bibnamefont{White}},
  \bibinfo{journal}{Phys. Rev. A} \textbf{\bibinfo{volume}{64}},
  \bibinfo{pages}{052312} (\bibinfo{year}{2001}).

\bibitem[{\citenamefont{Azzam}(McGraw-Hill, New York, 1995)}]{handbook2}
\bibinfo{author}{\bibfnamefont{R.~M.~A.} \bibnamefont{Azzam}},
  \bibinfo{journal}{in {\it Handbook of Optics}, 2nd Edition, Vol. II, Chap.
  27, edited by M. Bass, E. W. van Stryland, D. R. Williams and W. L. Wolfe}
  (\bibinfo{year}{McGraw-Hill, New York, 1995}).

\bibitem[{\citenamefont{Ambirajan and {Look Jr}}(1995{\natexlab{a}})}]{ambi1}
\bibinfo{author}{\bibfnamefont{A.}~\bibnamefont{Ambirajan}} \bibnamefont{and}
  \bibinfo{author}{\bibfnamefont{D.~C.} \bibnamefont{{Look Jr}}},
  \bibinfo{journal}{Opt. Eng.} \textbf{\bibinfo{volume}{34}},
  \bibinfo{pages}{(6), 1651} (\bibinfo{year}{1995}{\natexlab{a}}).

\bibitem[{\citenamefont{Ambirajan and {Look Jr}}(1995{\natexlab{b}})}]{ambi2}
\bibinfo{author}{\bibfnamefont{A.}~\bibnamefont{Ambirajan}} \bibnamefont{and}
  \bibinfo{author}{\bibfnamefont{D.~C.} \bibnamefont{{Look Jr}}},
  \bibinfo{journal}{Opt. Eng.} \textbf{\bibinfo{volume}{34}},
  \bibinfo{pages}{(6), 1656} (\bibinfo{year}{1995}{\natexlab{b}}).

\bibitem[{\citenamefont{Azzam and De}(2003)}]{doafdp}
\bibinfo{author}{\bibfnamefont{R.}~\bibnamefont{Azzam}} \bibnamefont{and}
  \bibinfo{author}{\bibfnamefont{A.}~\bibnamefont{De}}, \bibinfo{journal}{J.
  Opt. Soc. Am. A} \textbf{\bibinfo{volume}{20}}, \bibinfo{pages}{(5), 955}
  (\bibinfo{year}{2003}).

\bibitem[{\citenamefont{Sabatke et~al.}(2000)\citenamefont{Sabatke, Descour,
  Dereniak, Sweatt, Kemme, and Phipps}}]{sabatke1}
\bibinfo{author}{\bibfnamefont{D.~S.} \bibnamefont{Sabatke}},
  \bibinfo{author}{\bibfnamefont{M.~R.} \bibnamefont{Descour}},
  \bibinfo{author}{\bibfnamefont{E.~L.} \bibnamefont{Dereniak}},
  \bibinfo{author}{\bibfnamefont{W.~C.} \bibnamefont{Sweatt}},
  \bibinfo{author}{\bibfnamefont{S.~A.} \bibnamefont{Kemme}}, \bibnamefont{and}
  \bibinfo{author}{\bibfnamefont{G.~S.} \bibnamefont{Phipps}},
  \bibinfo{journal}{Opt. Lett.} \textbf{\bibinfo{volume}{25}},
  \bibinfo{pages}{(11), 802} (\bibinfo{year}{2000}).

\bibitem[{\citenamefont{Bru{\ss} et~al.}(2003)\citenamefont{Bru{\ss},
  Christandl, Ekert, Englert, Kaszlikowski, and Macchiavello}}]{tomo2}
\bibinfo{author}{\bibfnamefont{D.}~\bibnamefont{Bru{\ss}}},
  \bibinfo{author}{\bibfnamefont{M.}~\bibnamefont{Christandl}},
  \bibinfo{author}{\bibfnamefont{A.}~\bibnamefont{Ekert}},
  \bibinfo{author}{\bibfnamefont{B.-G.} \bibnamefont{Englert}},
  \bibinfo{author}{\bibfnamefont{D.}~\bibnamefont{Kaszlikowski}},
  \bibnamefont{and}
  \bibinfo{author}{\bibfnamefont{C.}~\bibnamefont{Macchiavello}},
  \bibinfo{journal}{Phys. Rev. Lett.} \textbf{\bibinfo{volume}{91}},
  \bibinfo{pages}{097901} (\bibinfo{year}{2003}).

\bibitem[{\citenamefont{Liang et~al.}(2003)\citenamefont{Liang, Kaszlikowski,
  Englert, Kwek, and Oh}}]{tomo1}
\bibinfo{author}{\bibfnamefont{Y.~C.} \bibnamefont{Liang}},
  \bibinfo{author}{\bibfnamefont{D.}~\bibnamefont{Kaszlikowski}},
  \bibinfo{author}{\bibfnamefont{B.-G.} \bibnamefont{Englert}},
  \bibinfo{author}{\bibfnamefont{L.~C.} \bibnamefont{Kwek}}, \bibnamefont{and}
  \bibinfo{author}{\bibfnamefont{C.~H.} \bibnamefont{Oh}},
  \bibinfo{journal}{Phys. Rev. A} \textbf{\bibinfo{volume}{68}},
  \bibinfo{pages}{022324} (\bibinfo{year}{2003}).

\bibitem[{\citenamefont{Englert
  et~al.}(2005{\natexlab{a}})\citenamefont{Englert, Kaszlikowski, Ng, Chua,
  \u{R}eh\'{a}\u{c}ek, and Anders}}]{sprot}
\bibinfo{author}{\bibfnamefont{B.-G.} \bibnamefont{Englert}},
  \bibinfo{author}{\bibfnamefont{D.}~\bibnamefont{Kaszlikowski}},
  \bibinfo{author}{\bibfnamefont{H.~K.} \bibnamefont{Ng}},
  \bibinfo{author}{\bibfnamefont{W.~K.} \bibnamefont{Chua}},
  \bibinfo{author}{\bibfnamefont{J.}~\bibnamefont{\u{R}eh\'{a}\u{c}ek}},
  \bibnamefont{and} \bibinfo{author}{\bibfnamefont{J.}~\bibnamefont{Anders}},
  \bibinfo{journal}{arXiv:quant-ph/0412075}
  (\bibinfo{year}{2005}{\natexlab{a}}).

\bibitem[{\citenamefont{Andersson et~al.}(2005)\citenamefont{Andersson,
  Barnett, and Aspect}}]{aspect}
\bibinfo{author}{\bibfnamefont{E.}~\bibnamefont{Andersson}},
  \bibinfo{author}{\bibfnamefont{S.~M.} \bibnamefont{Barnett}},
  \bibnamefont{and} \bibinfo{author}{\bibfnamefont{A.}~\bibnamefont{Aspect}},
  \bibinfo{journal}{Phys. Rev. A} \textbf{\bibinfo{volume}{72}},
  \bibinfo{pages}{042104} (\bibinfo{year}{2005}).

\bibitem[{\citenamefont{Azzam et~al.}(1988)\citenamefont{Azzam, Elminyawi, and
  El-Saba}}]{anafdp}
\bibinfo{author}{\bibfnamefont{R.}~\bibnamefont{Azzam}},
  \bibinfo{author}{\bibfnamefont{I.}~\bibnamefont{Elminyawi}},
  \bibnamefont{and} \bibinfo{author}{\bibfnamefont{A.}~\bibnamefont{El-Saba}},
  \bibinfo{journal}{Opt. Soc. Am.} \textbf{\bibinfo{volume}{5}},
  \bibinfo{pages}{(5), 681} (\bibinfo{year}{1988}).

\bibitem[{blo()}]{bloch}
\bibinfo{journal}{In the language of spin-$\frac{1}{2}$ systems the reduced
  Stokes vector is the Pauli vector and the Poincare sphere is called the Bloch
  sphere.}

\bibitem[{\citenamefont{Kraus}(Springer-Verlag, Berlin, 1983)}]{kraus}
\bibinfo{author}{\bibfnamefont{K.}~\bibnamefont{Kraus}}, \bibinfo{journal}{{\it
  States, Effects and Operations. Fundamental Notions of Quantum Theory},
  Lecture Notes in Physics Vol. 190}  (\bibinfo{year}{Springer-Verlag, Berlin,
  1983}).

\bibitem[{\citenamefont{Englert
  et~al.}(2005{\natexlab{b}})\citenamefont{Englert, Tin, Goh, and Ng}}]{sle}
\bibinfo{author}{\bibfnamefont{B.-G.} \bibnamefont{Englert}},
  \bibinfo{author}{\bibfnamefont{K.~M.} \bibnamefont{Tin}},
  \bibinfo{author}{\bibfnamefont{C.~G.} \bibnamefont{Goh}}, \bibnamefont{and}
  \bibinfo{author}{\bibfnamefont{H.~K.} \bibnamefont{Ng}},
  \bibinfo{journal}{Laser Phys.} \textbf{\bibinfo{volume}{15}},
  \bibinfo{pages}{7} (\bibinfo{year}{2005}{\natexlab{b}}).

\bibitem[{\citenamefont{O'Neill}(1991)}]{statopt}
\bibinfo{author}{\bibfnamefont{E.~L.} \bibnamefont{O'Neill}},
  \bibinfo{journal}{Chap. 9 in {\it Introduction to Statistical Optics}, Dover
  Publications Inc., N.Y.}  (\bibinfo{year}{1991}).

\bibitem[{\citenamefont{Azzam and Sudrajat}(2005)}]{doafdpbs}
\bibinfo{author}{\bibfnamefont{R.~M.~A.} \bibnamefont{Azzam}} \bibnamefont{and}
  \bibinfo{author}{\bibfnamefont{F.~F.} \bibnamefont{Sudrajat}},
  \bibinfo{journal}{Applied Optics} \textbf{\bibinfo{volume}{44}},
  \bibinfo{pages}{(2), 190} (\bibinfo{year}{2005}).

\bibitem[{\citenamefont{Kurtsiefer et~al.}(2001)\citenamefont{Kurtsiefer,
  Oberparleiter, and Weinfurter}}]{brsrc}
\bibinfo{author}{\bibfnamefont{C.}~\bibnamefont{Kurtsiefer}},
  \bibinfo{author}{\bibfnamefont{M.}~\bibnamefont{Oberparleiter}},
  \bibnamefont{and}
  \bibinfo{author}{\bibfnamefont{H.}~\bibnamefont{Weinfurter}},
  \bibinfo{journal}{Phys. Rev. A} \textbf{\bibinfo{volume}{64}},
  \bibinfo{pages}{023802} (\bibinfo{year}{2001}).

\bibitem[{\citenamefont{Uhlmann}(1976)}]{uhlmann}
\bibinfo{author}{\bibfnamefont{A.}~\bibnamefont{Uhlmann}},
  \bibinfo{journal}{Rep. Math. Phys.} \textbf{\bibinfo{volume}{9}},
  \bibinfo{pages}{273} (\bibinfo{year}{1976}).

\bibitem[{\citenamefont{Jozsa}(1994)}]{jozsa}
\bibinfo{author}{\bibfnamefont{R.}~\bibnamefont{Jozsa}}, \bibinfo{journal}{J.
  Mod. Optics} \textbf{\bibinfo{volume}{41}}, \bibinfo{pages}{2315}
  (\bibinfo{year}{1994}).

\bibitem[{\citenamefont{Gaertner et~al.}(2005)\citenamefont{Gaertner,
  Weinfurter, and Kurtsiefer}}]{multichannel}
\bibinfo{author}{\bibfnamefont{S.}~\bibnamefont{Gaertner}},
  \bibinfo{author}{\bibfnamefont{H.}~\bibnamefont{Weinfurter}},
  \bibnamefont{and}
  \bibinfo{author}{\bibfnamefont{C.}~\bibnamefont{Kurtsiefer}},
  \bibinfo{journal}{Rev. Sci. Inst.} \textbf{\bibinfo{volume}{76}},
  \bibinfo{pages}{123108} (\bibinfo{year}{2005}).

\end{thebibliography}
\end{document}